# Comparing Standard Distribution and Its Tsallis Form of Transverse Momenta in High Energy Collisions

Rui-Fang Si[1,2], Hui-Ling Li[3], Fu-Hu Liu[1]

[1]*Institute of Theoretical Physics & State Key Laboratory of Quantum Optics and Quantum Optics Devices, Shanxi University, Taiyuan, Shanxi 030006, China*

[2]*Department of Mathematics and Science, Fenyang Normal Campus Lvliang College, Fenyang, Shanxi 032300, China*

[3]*Institute of Modern Physics, Shanxi Normal University, Linfen, Shanxi 041004, China*

Correspondence should be addressed to Hui-Ling Li; lihuilingxyx@sina.com and Fu-Hu Liu; fuhuliu@163.com

**Abstract**：In this paper, the experimental (simulated) transverse momentum spectra of negatively charged pions produced at mid-rapidity in central nucleus-nucleus collisions at the Heavy Ion Synchrotron (SIS), Relativistic Heavy Ion Collider (RHIC), and Large Hadron Collider (LHC) energies obtained by different collaborations are selected by us to investigate, where a few simulated data are taken from the results of FOPI Collaboration who uses the IQMD transport code based on Quantum Molecular Dynamics. A two-component standard distribution and the Tsallis form of standard distribution are used to fit these data in the framework of a multisource thermal model. The excitation functions of main parameters in the two distributions are analyzed. In particular, the effective temperatures extracted from the two-component standard distribution and the Tsallis form of standard distribution are obtained, and the relation between the two types of effective temperatures is studied.

## 1. Introduction

High energy heavy ion (nucleus-nucleus) collisions are an important method to simulate and study the big bang in the early universe, properties of new matter created in extreme conditions, accompanying phenomena in the creation, and physics mechanisms of the creation. Some models based on the quantum chromodynamics (QCD) and/or thermal and statistical methods can be used to analyze the equation of state (EoS) at finite temperature and density, properties of chemical and kinetic freeze-outs in collision process, distribution laws of different particles in final state, and universality of hadroproduction in different systems [1-5]. The properties of nuclear matter and its phase transition to quark-gluon plasma (QGP) at high temperature and density can be obtained. With the developments in the methodologies of experimental techniques and theoretical studies, the collision energy per nucleon pair in the center-of-mass system increases from high energy range which has a few to several hundred GeV to ultrahigh energy range which has presently a few to over ten TeV.

The temperature and density described the EoS show that the new matter created in high and ultrahigh energy ranges is not similar to the ideal gas-like state of quarks and gluons expected by early theoretical models. Instead, the effects of strong dynamical coupling, long-range interactions, local memory, and others appear in the interior of interacting system. The rapid evolution of interacting system and the indirect measurements of some observable quantities result in that one can use the statistical method to study the distribution properties of some observable quantities such as (pseudo)rapidity, (transverse) momentum, (transverse) energy, azimuthal angle, elliptic flow, multiplicity, and others of final-state fragments and particles [1-5]. Thus, some quantitative or qualitative results related to the properties of interacting system and particle production can be observed.

As the quantities which can be early measured in experiments, i.e. the so-called "the first day" measureable quantities, the rapidity and transverse momentum distributions attract wide attentions due to their carry-overs on the information of longitudinal extension and transverse expansion of the emission source in interacting system. With the increasing collision energy, the rapidity distribution range extends from a few rapidity units to over ten rapidity units, and the transverse momentum distribution range increases from 0 until a few GeV/$c$ to 0 until over hundred GeV/$c$. Different functions and methods are used by different researchers to describe rapidity and transverse momentum distributions as well as other distributions which can be measured in experiments [1-5]. Based on a multi-source thermal model [6-9], the rapidity and transverse momentum distributions obtained in experiments at different collision energies are studied by us in terms of two-cylinder, Rayleigh, Boltzmann, Tsallis, and other distributions. In particular, comparing with rapidity distribution, transverse momentum distribution contains more abundant information and attracts wider attentions. Although one has Monte Carlo and other indirect methods to describe transverse momentum distributions, analytical functions are more expected to use.

Because of the same transverse momentum distribution being described by different functions to obtain values of different parameters, possible relations existed among different parameters can be studied. In this paper, based on the multi-source thermal model [6-9], the standard distribution (Boltzmann, Fermi-Dirac, and Bose-Einstein distributions) and its Tsallis form are used to describe the transverse momentum distribution of final-state particles produced in high energy nucleus-nucleus collisions. The excitation functions of effective temperatures obtained by the two distributions are extracted and the relation between the two effective temperatures is studied.

The rest part of this paper is structured as followings. A briefly description of the model and method is presented in section 2. Results on comparisons with experimental (simulated) data and discussion are given in section 3. Finally, we summarize our main observations and conclusions in section 4.

**2. The model and method**

According to the multi-source model [6-9], a few emission sources of produced particles are assumed to form in interacting system due to different reaction mechanisms and/or data examples. For each emission source, the thermal model or other similar models and distributions can be used to perform calculation on the production of particles. The potential models include [10], but are not limited to, ideal gas-like model, ideal hydrodynamic model, viscous hydrodynamic model, and others. In these models, the relativistic effect has to be particularly considered, and the quantum effect can be usually neglected. If we study in detail the interacting system and final-state particles, both the relativistic and quantum effects have to be considered.

In the middle stage of collision process, the interacting system and emission sources in it can be regarded as to stay at the hydrodynamic state. After the stage of chemical freeze-out, in particular after the stage of kinetic freeze-out, the interacting system and emission sources in it should stay at the gas-like state. Otherwise, it is difficult to understand the kinetic information of singular particle measured in experiments. As for what and why had happen during the phase transition from the liquid-like state at the middle stage to the gas-like state at the final stage is beyond the focus of the present work. We shall not discuss this issue here.

According to the ideal gas model with the relativistic and quantum effects, the particle spectra can be described by the standard distribution. The number of particles is [11]

$$N = \frac{gV}{(2\pi)^3} \int d^3p \left[ \exp\left( \pm \frac{E-\mu}{T_S} \right) + S \right]^{-1}, \qquad (1)$$

where $g$ is the degeneracy factor, $V$ is the volume, $p$ is the momentum, $E = \sqrt{p^2 + m_0^2}$ is the energy, $m_0$ is the rest mass, $\mu$ is the chemical potential, $T_S$ is the effective temperature; $S = 0$, $+1$, and $-1$ correspond to the Boltzmann, Fermi-Dirac, and Bose-Einstein statistics respectively; $E - \mu > 0$ corresponds to plus $+$, and $E - \mu \leq 0$ corresponds to minus $-$. The invariant momentum distribution of particles is

$$E \frac{d^3N}{dp^3} = \frac{gV}{(2\pi)^3} E \left[ \exp\left( \pm \frac{E-\mu}{T_S} \right) + S \right]^{-1}. \qquad (2)$$

The normalized probability density distribution of particle momenta can be written as

$$f_p(p) = \frac{1}{N}\frac{dN}{dp} = C_S p^2 \left[\exp\left(\pm\frac{\sqrt{p^2+m_0^2}-\mu}{T_S}\right)+S\right]^{-1}, \tag{3}$$

where $C_S$ is the normalized constant in the standard probability density distribution of momenta. It is related to the selection of parameters.

The normalized joint probability density distribution of particle rapidities and transverse momenta is

$$\begin{aligned}f_{y,p_T}(y,p_T) &= \frac{1}{N}\frac{d^2N}{dydp_T} \\ &= C_S p_T \sqrt{p_T^2+m_0^2}\cosh y\left[\exp\left(\pm\frac{\sqrt{p_T^2+m_0^2}\cosh y-\mu}{T_S}\right)+S\right]^{-1},\end{aligned} \tag{4}$$

where $\sqrt{p_T^2+m_0^2}\cosh y-\mu>0$ corresponds to plus +, and $\sqrt{p_T^2+m_0^2}\cosh y-\mu\leq 0$ corresponds to minus $-$. The normalized probability density distribution of particle rapidities is then written to be

$$\begin{aligned}f_y(y) &= \frac{1}{N}\frac{dN}{dy} \\ &= C_S \cosh y \int_0^{p_{T\max}} p_T\sqrt{p_T^2+m_0^2}\left[\exp\left(\pm\frac{\sqrt{p_T^2+m_0^2}\cosh y-\mu}{T_S}\right)+S\right]^{-1}dp_T,\end{aligned} \tag{5}$$

where $p_{T\max}$ denotes the maximum transverse momentum. This rapidity distribution is only for an emission source. In the case of considering multiple sources, we have to consider sources distribution in the rapidity space [2-4, 12-16]. This issue is beyond the focus of the present work, and we shall not discuss it anymore. The normalized probability density distribution of particle transverse momenta is written to be

$$\begin{aligned}f_{p_T}(p_T) &= \frac{1}{N}\frac{dN}{dp_T} \\ &= C_S p_T \sqrt{p_T^2+m_0^2}\int_{y_{\min}}^{y_{\max}}\cosh y\left[\exp\left(\pm\frac{\sqrt{p_T^2+m_0^2}\cosh y-\mu}{T_S}\right)+S\right]^{-1}dy,\end{aligned} \tag{6}$$

where $y_{\max}$ and $y_{\min}$ denote the maximum and minimum rapidities respectively.

It should be noted that, in the above formulas, although the same symbol $C_S$ is used to represent the normalized constants in different formulas, these constants may be different each other. In the case of considering multi-source emission, we have to use the multi-component distribution to describe the transverse momentum distribution of final-state particles. If $n_0$ emission sources are considered, we have

$$f_{p_T}(p_T) = \frac{1}{N}\frac{dN}{dp_T}$$

$$= \sum_{i=1}^{n_0} k_{Si} C_{Si} p_T \sqrt{p_T^2 + m_0^2} \int_{y_{\min}}^{y_{\max}} \cosh y \left[ \exp\left( \pm \frac{\sqrt{p_T^2 + m_0^2}\cosh y - \mu}{T_{Si}} \right) + S \right]^{-1} dy, \quad (7)$$

where $C_{Si}$ denotes the normalized constant for the $i$th component in $n_0$ components, $k_{Si}$ denotes the contribution fraction of the $i$th component in final-state distribution, and $T_{Si}$ denotes the effective temperature corresponding to the $i$th component. There are temperature fluctuations among different components. In the case of considering multi-source emission, we have the effective temperature of interacting system to be $T_S = \sum_i k_{Si} T_{Si}$. Generally, two or three emission sources are enough to describe the experimental data obtained in soft excitation process. That is $i = 2$ or 3 in most cases.

If we consider the Tsallis form of standard distribution, the number of particles is [11, 17]

$$N = \frac{gV}{(2\pi)^3} \int d^3 p \left\{ \left[ 1 \pm \frac{q-1}{T_T}(E-\mu) \right]^{\pm 1/(q-1)} + S \right\}^{-1}, \quad (8)$$

where $q$ is an entropy index which characterizes the departing degree of the interacting system from the equilibrium state. Generally, we have $q > 1$; If $q = 1$, the system stays in the equilibrium state. $T_T$ is the effective temperature. Other symbols have the same meanings as Eq. (1). The invariant momentum distribution of particles is

$$E\frac{d^3 N}{dp^3} = \frac{gV}{(2\pi)^3} E \left\{ \left[ 1 \pm \frac{q-1}{T_T}(E-\mu) \right]^{\pm 1/(q-1)} + S \right\}^{-1}. \quad (9)$$

The normalized probability density distribution of particle momenta is

$$f_p(p) = \frac{1}{N}\frac{dN}{dp} = C_T p^2 \left\{ \left[ 1 \pm \frac{q-1}{T_T}\left(\sqrt{p^2 + m_0^2} - \mu\right) \right]^{\pm 1/(q-1)} + S \right\}^{-1}. \quad (10)$$

The normalized joint probability density distribution of particle rapidities and transverse momenta is

$$f_{y,p_T}(y, p_T) = \frac{1}{N}\frac{d^2 N}{dy dp_T}$$

$$= C_T p_T \sqrt{p_T^2 + m_0^2} \cosh y \left\{ \left[ 1 \pm \frac{q-1}{T_T}\left(\sqrt{p_T^2 + m_0^2}\cosh y - \mu\right) \right]^{\pm 1/(q-1)} + S \right\}^{-1}. \quad (11)$$

Then, the normalized probability density distribution of particle rapidities is

$$f_y(y) = \frac{1}{N}\frac{dN}{dy}$$

$$= C_T \cosh y \int_0^{p_{T\max}} p_T \sqrt{p_T^2 + m_0^2} \left\{ \left[1 \pm \frac{q-1}{T_T}\left(\sqrt{p_T^2 + m_0^2}\cosh y - \mu\right)\right]^{\pm 1/(q-1)} + S \right\}^{-1} dp_T. \quad (12)$$

The normalized probability density distribution of particle transverse momenta is

$$f_{p_T}(p_T) = \frac{1}{N}\frac{dN}{dp_T}$$

$$= C_T p_T \sqrt{p_T^2 + m_0^2} \int_{y_{\min}}^{y_{\max}} \cosh y \left\{ \left[1 \pm \frac{q-1}{T_T}\left(\sqrt{p_T^2 + m_0^2}\cosh y - \mu\right)\right]^{\pm 1/(q-1)} + S \right\}^{-1} dy. \quad (13)$$

In the above formulas, although the same symbol $C_T$ is used to represent the normalized constants in different formulas, these constants may be different each other. As discussed in ref. [17], the Tsallis form has at least four types of function representations, though we choose only one that contains $p_T$ after $C_T$ and the index $1/(q-1)$. We do not need to consider a multi-source for the Tsallis form due to it covering a two- or three-component standard distribution, and the two- or three-component standard distribution describes well the transverse momentum spectrum of particles produced in soft excitation process.

It should be noted again that, the above multi-component (two- or three-component) standard distribution and the Tsallis form of standard distribution can describe only the transverse momentum spectrum of particles produced in soft excitation process. The transverse momentum spectrum produced in soft excitation process covers a narrow range. For the transverse momentum spectrum covering a wide range, we have to consider the contribution of hard scattering process. According to the QCD calculus [18-20], we have an inverse power-law

$$f_H(p_T) = \frac{1}{N}\frac{dN}{dp_T} = Ap_T\left(1 + \frac{p_T}{p_0}\right)^{-n} \quad (14)$$

to describe the transverse momentum spectrum produced in hard scattering process, where $p_0$ and $n$ are free parameters, and $A$ is the normalized constant which is related to the free parameters. It is obviously that a two-component function is needed for a wide transverse momentum spectrum. The first component is the multi-component (two- or three-component) standard distribution or Tsallis form which describe the soft process, and the second component is the inverse power-law which describes the hard process. The application of the inverse power-law is beyond the focus of the present work. We shall not discuss it anymore.

In the above discussions, to obtain chemical potential of a given particle, the chemical freeze-out temperature $T_{ch}$ of the emission source is needed to know firstly. In the case of assuming the same chemical freeze-out moment, the emission source has the sole $T_{ch}$. According to refs. [21, 22], there is a relation among $T_{ch}$, the yield $n_1$ and mass $m_1$ of the first particle, the yield $n_2$ and mass $m_2$ of the second particle, and the ratio $n_{12} = n_1/n_2$. We have

$$n_{12} = \frac{n_1}{n_2} = \frac{\exp(m_2/T_{ch}) + S_2}{\exp(m_1/T_{ch}) + S_1}, \qquad (15)$$

where $S_1(S_2) = \pm 1$ denote fermion and boson respectively. If the fermion and boson are not needed to distinguish, we have $S = 0$. This results in a simple expression for Eq. (15), that is $n_{12} = n_1/n_2 \approx \exp(-m_1/T_{ch})/\exp(-m_2/T_{ch})$.

In the framework of a statistical thermal model of non-interacting gas particles with the assumption of standard Maxwell-Boltzmann statistics, there is an empirical expression for the chemical freeze-out temperature [23-26],

$$T_{ch} = \frac{0.164}{1 + \exp\left[2.60 - \ln\left(\sqrt{s_{NN}}\right)/0.45\right]}, \qquad (16)$$

where $\sqrt{s_{NN}}$ denotes the energy per nucleon pair in the center-of-mass system. Both the units of $T_{ch}$ and $\sqrt{s_{NN}}$ are in GeV. The limiting value of $T_{ch}$ is 0.164 GeV.

In the framework of a thermal model with standard distribution, the chemical potentials of some particles can be obtained from the ratios of negatively to positively charged particles. According to [27], we have

$$\frac{\bar{p}}{p} = \exp\left(-\frac{2\mu_p}{T_{ch}}\right) \equiv k_p, \qquad (17)$$

$$\frac{K^-}{K^+} = \exp\left(-\frac{2\mu_K}{T_{ch}}\right) \equiv k_K, \qquad (18)$$

$$\frac{\pi^-}{\pi^+} = \exp\left(-\frac{2\mu_\pi}{T_{ch}}\right) \equiv k_\pi, \qquad (19)$$

where the symbol of a given particle is used for its yield for the purpose of simplicity. Further, the chemical potentials of the mentioned particles are

$$\mu_p = -\frac{1}{2}T_{ch} \cdot \ln(k_p), \qquad (20)$$

$$\mu_K = -\frac{1}{2}T_{ch} \cdot \ln(k_K), \qquad (21)$$

$$\mu_\pi = -\frac{1}{2} T_{ch} \cdot \ln(k_\pi). \tag{22}$$

Empirically, the chemical potential for baryon is [23-26]

$$\mu_B = \frac{1.303}{1 + 0.286\sqrt{s_{NN}}} \tag{23}$$

which is also obtained in the framework of a statistical thermal model of non-interacting gas particles with the assumption of standard Maxwell-Boltzmann statistics, where both the units of $\mu_B$ and $\sqrt{s_{NN}}$ are in GeV.

We would like to point out that Eqs. (16) and (23) should be modified in the framework of a generalized non-extensive statistics when we use the Tsallis form of standard distribution. At the same time, Eqs. (17)-(22) should be generalized within an analysis with the Tsallis form. To modify Eqs. (16)-(23) is beyond our focus and ability. We shall not discuss these modifications here. Instead, as an approximate treatment, we use $T_{ch}$ and $\mu_\pi$ obtained within an analysis with the standard distribution as those within the Tsallis form. In fact, the absolute value of $\mu_\pi$ is very small, and its effect on the transverse momentum spectra can be neglected. Therefore, this approximate treatment is acceptable.

It should be noted once more that, as mentioned in the above discussions, what we extract from the multi-component standard distribution or the Tsallis form of standard distribution is the effective temperature, but not the real temperature of emission source. Generally, the transverse momentum spectrum contains both the contributions of thermal motion and flow effect. The real temperature is only a reflection of purely thermal motion, and the flow effect should not be included in it. As for the methods to obtain the real temperature by disengaging the contributions of thermal motion and flow effect, we can use the blast-wave model based on the Boltzmann distribution [28-30], the blast-wave model based on the Tsallis distribution [31], the improved Tsallis distribution [32, 33], some alternative methods [21, 29, 34-36], and others [37-40]. These methods themselves are beyond the focus of the present work. We shall not discuss them anymore.

## 3. Results and discussion

The transverse momentum spectra of negatively charged pions produced in mid-rapidity range in $\sqrt{s_{NN}}$ = 2.24 and 2.52 GeV central gold-gold (Au-Au) collisions [41] measured (simulated) by the FOPI Collaboration at the Heavy Ion Synchrotron (SIS), 11.5 [42], 62.4, 130, and 200 GeV central Au-Au collisions [29] measured by the STAR Collaboration at the Relativistic Heavy Ion Collider (RHIC), 22.5 GeV central copper-copper (Cu-Cu) [43] and 200 GeV central Au-Au collisions [27] measured by the PHENIX Collaboration at the RHIC, and 2.76 TeV central lead-lead (Pb-Pb) collisions [44] measured by the ALICE Collaboration at the Large Hadron Collider (LHC) are selected to investigate. Among them, the results of FOPI Collaboration are given in Figure 1 with the simulated data (the last eight circles) of the IQMD transport code [46] which is based on Quantum Molecular Dynamics [47]; To avoid huddle, most results of the STAR Collaboration are given in Figure 2, and the results corresponding to 11.5 GeV are given in Figure 3; The results of PHENIX and ALICE Collaborations are given in Figures 3 and 4 respectively. In each figure, the symbols represent the experimental (simulated) data scaled by different amounts in some cases. The collision energy and type, centrality and mid-rapidity ranges, and scaled amount if not 1 are marked in the panel. The dashed and solid curves denote the results fitted by the two-component standard distribution and the Tsallis form of standard distribution. The values of parameters, $\chi^2$, and degree of freedom ($dof$) are listed in Table 1 ordered by the energy from low to high. In particular, $T_S = k_{S1}T_{S1} + (1-k_{S1})T_{S2}$ is the average weighted by the fractions of different components, $\mu_\pi$ is obtained by Eqs. (16) and (22), and the values of $k_\pi$ in Eq. (22) at different energies are obtained from ref. [45]. As a preliminary result, the values of $\mu_\pi$ for the first and second standard distribution and the Tsallis form are assumed to be the same. In the fitting, the method of least square is used to obtain the best parameter values. One can see that the two-component standard distribution and the Tsallis form of standard distribution describe approximately the transverse momentum spectra of negatively charged pions produced in central nucleus-nucleus collisions in the energy range from SIS to LHC.

To study the excitation functions of free parameters, i.e. the dependences of free parameters on collision energy, the relations $T_{S1} - \ln\sqrt{s_{NN}}$ ($T_{S2} - \ln\sqrt{s_{NN}}$), $T_S - \ln\sqrt{s_{NN}}$ ($T_T - \ln\sqrt{s_{NN}}$), $k_{S1} - \ln\sqrt{s_{NN}}$ and $q - \ln\sqrt{s_{NN}}$ are presented in Figures 5-8, respectively. The symbols and error bars in the figures denote the values of free parameters and their errors. Both the values of free parameters and their errors are taken from Table 1. The lines in Figures 5 and 6 are obtained by the method of least square. These lines can be described by linear functions $T_{S1,S2,S,T} = a\ln\sqrt{s_{NN}} + b$, where the slope $a$ and intercept $b$ are listed in Table 1 and the unit of $\sqrt{s_{NN}}$ is in GeV. One can see that the four effective temperatures $T_{S1}$, $T_{S2}$, $T_S$, and $T_T$ increase linearly with increase of $\ln\sqrt{s_{NN}}$. In particular, the relation between $T_S$ and $T_T$ can be obtained to be $T_S = (2.500 \pm 0.170)T_T + (-0.040 \pm 0.013)$ due to Table 1, which shows a linear relation between $T_S$ and $T_T$. With increase of $\sqrt{s_{NN}}$, $k_{S1}$ has a minimum at about 10 GeV, and $q$ increases primitively and saturates at about 10 GeV.

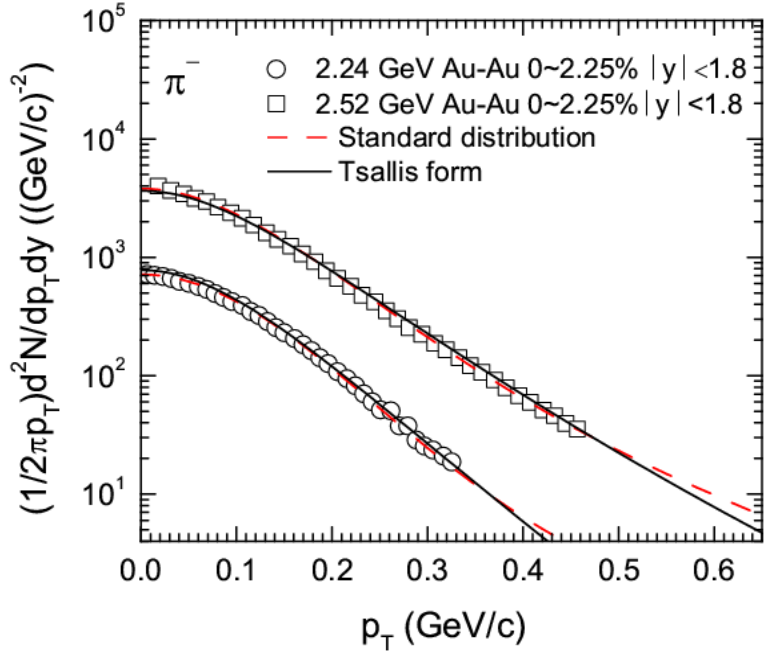

Figure 1: Transverse momentum spectra of $\pi^-$ produced in central Au-Au collisions at $\sqrt{s_{NN}} = 2.24$ (circles) and 2.52 GeV (squares). The symbols represent the experimental data of the FOPI Collaboration [41] measured in mid-rapidity range and the last eight circles represent the simulated data of the IQMD transport code [46] which is based on Quantum Molecular Dynamics [47]. The statistical errors are smaller than the size of symbols. The dashed and solid curves are the results fitted by the two-component standard distribution and the Tsallis form of standard distribution respectively.

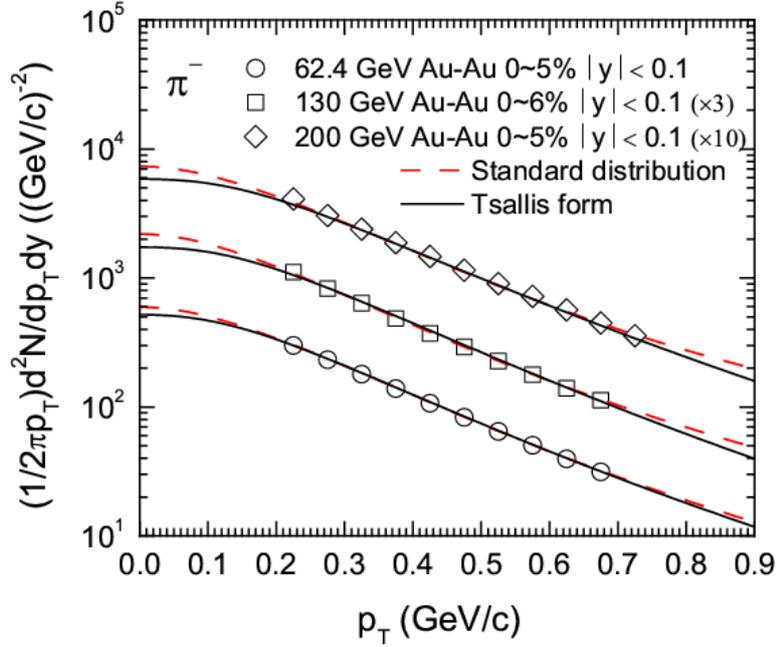

Figure 2: Same as Figure 1, but for $\sqrt{s_{NN}} = 62.4$ (circles), 130 (squares), and 200 GeV (rhombuses). The symbols represent the experimental data of the STAR Collaboration with errors to be the root quadratic sum of statistical and systematic errors [29].

Our results show some interesting features. Actually, one could as well say there is no difference in the particle production in central nucleus-nucleus collisions from a few GeV to a few TeV. This in some sense echoes recent studies of Sarkisyan et al. [1, 48]. In addition, our recent study shows that the same or similar fits to be good for proton-proton collisions [40], though the parameter values in proton-proton collisions are closer to those in peripheral nucleus-nucleus collisions when comparing with central nucleus-nucleus collisions. This suggests universality in particle production, as it is obtained in recent and previous studies of Sarkisyan et al. [1, 12, 48-50], but now for transverse momentum distribution as well. On the other hand, the multiplicity and transverse momentum distributions observed in different data samples can be uniformly fitted by multi-component Erlang distribution [8, 51, 52], which also show the universality in particle production. Indeed, the universality in particle production exists not only in mean multiplicity and pseudorapidity density but also in multiplicity and transverse momentum distributions in some conditions.

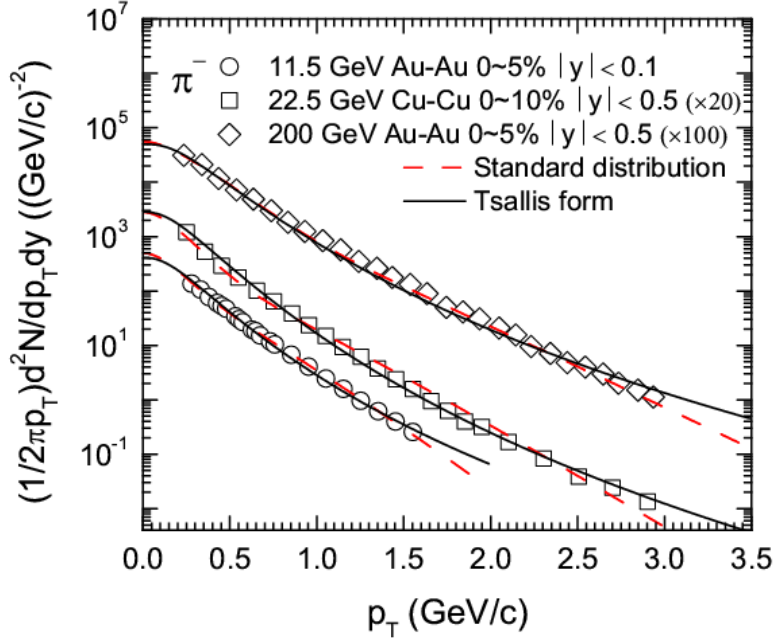

Figure 3: Same as Figure 1, but for central Au-Au collisions at $\sqrt{s_{NN}} = 11.5$ GeV (circles), central Cu-Cu collisions at $\sqrt{s_{NN}} = 22.5$ GeV (squares), and central Au-Au collisions at $\sqrt{s_{NN}} = 200$ GeV (rhombuses). The symbols represent the experimental data of the STAR Collaboration with the statistical and systematic errors added in quadrature (circles) [42], and the PHENIX Collaborations with statistical errors only (squares [43] and rhombuses [27]).

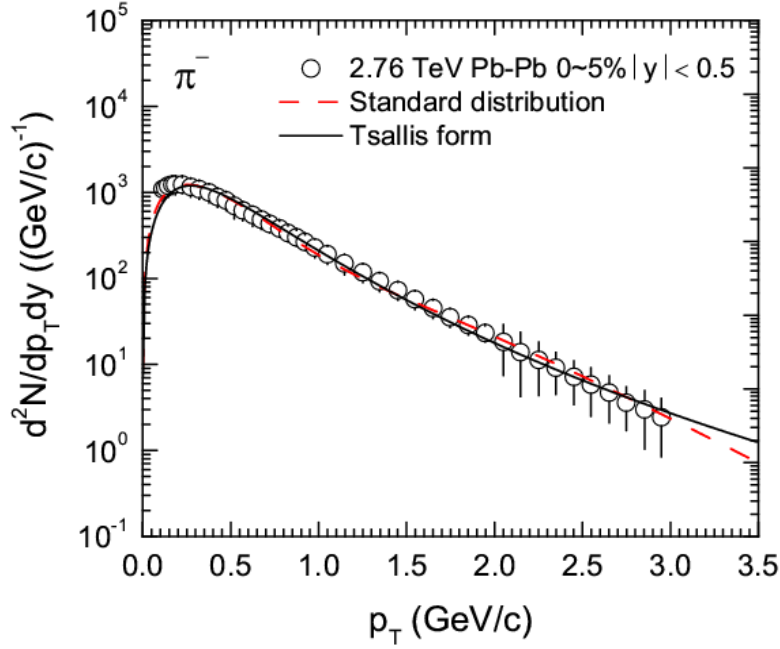

Figure 4: Same as Figure 1, but for central Pb-Pb collisions at $\sqrt{s_{NN}} = 2.76$ TeV. The symbols represent the experimental data of the ALICE Collaborations and errors are the root quadratic sum of statistical and systematic errors [44].

Table 1: Values of free parameters ($T_{S1}$, $T_{S2}$, and $k_{S1}$) for the two-component standard distribution, free parameters ($T_T$ and $q$) for the Tsallis form of standard distribution, leading-out parameters ($T_S$ and $\mu_\pi$), $\chi^2$, and $dof$ corresponding to the fits in Figures 1-4, as well as values of slope $a$ and intercept $b$ corresponding to the fits of linear functions $T_{S1,S2,S,T} = a\ln\sqrt{s_{NN}} + b$ in Figures 5 and 6. As a preliminary result, the values of $\mu_\pi$ for the first and second standard distribution and the Tsallis form are assumed to be the same. The units of effective temperatures and chemical potential are in GeV.

|  | $T_{S1}$ | $T_{S2}$ | $k_{S1}$ | $T_S$ | $\mu_\pi$ | $\chi^2/dof$ | $T_T$ | $q$ | $\chi^2/dof$ |
|---|---|---|---|---|---|---|---|---|---|
| 2.24 GeV Au-Au (FOPI Collab.) | 0.038±0.004 | 0.100±0.050 | 0.825±0.100 | 0.049±0.012 | −(0.021±0.002) | 14.880/31 | 0.039±0.005 | 1.042±0.010 | 24.043/32 |
| 2.52 GeV Au-Au (FOPI Collab.) | 0.054±0.007 | 0.126±0.012 | 0.810±0.050 | 0.068±0.008 | −(0.019±0.003) | 11.004/42 | 0.040±0.003 | 1.072±0.008 | 16.856/43 |
| 11.5 GeV Au-Au (STAR Collab.) | 0.076±0.010 | 0.175±0.009 | 0.432±0.070 | 0.132±0.009 | −(0.005±0.007) | 1.350/18 | 0.067±0.008 | 1.106±0.008 | 2.850/19 |
| 22.5 GeV Cu-Cu (PHENIX Collab.) | 0.088±0.025 | 0.219±0.008 | 0.620±0.085 | 0.138±0.019 | −(0.003±0.011) | 4.123/19 | 0.075±0.005 | 1.089±0.004 | 1.360/20 |
| 62.4 GeV Au-Au (STAR Collab.) | 0.106±0.012 | 0.242±0.029 | 0.620±0.009 | 0.158±0.019 | −(0.002±0.014) | 2.130/6 | 0.070±0.005 | 1.131±0.011 | 1.561/7 |
| 130 GeV Au-Au (STAR Collab.) | 0.102±0.012 | 0.250±0.060 | 0.600±0.085 | 0.161±0.031 | −(0.001±0.015) | 6.246/6 | 0.076±0.005 | 1.117±0.016 | 42.448/7 |
| 200 GeV Au-Au (STAR Collab.) | 0.112±0.010 | 0.310±0.085 | 0.671±0.080 | 0.177±0.035 | −(0.002±0.018) | 10.241/7 | 0.078±0.006 | 1.120±0.015 | 10.736/8 |
| 200 GeV Au-Au (PHENIX Collab.) | 0.134±0.020 | 0.270±0.010 | 0.761±0.092 | 0.167±0.018 | −(0.002±0.018) | 3.048/24 | 0.089±0.004 | 1.096±0.005 | 7.650/25 |
| 2.76 TeV Pb-Pb (ALICE Collab.) | 0.145±0.010 | 0.345±0.011 | 0.761±0.040 | 0.193±0.010 | −(0.001±0.011) | 6.475/37 | 0.097±0.005 | 1.112±0.005 | 9.766/38 |
| $a$ | 0.015±0.012 | 0.034±0.022 |  | 0.020±0.023 |  |  | 0.008±0.008 |  |  |
| $b$ | 0.035±0.004 | 0.088±0.007 |  | 0.067±0.008 |  |  | 0.039±0.003 |  |  |

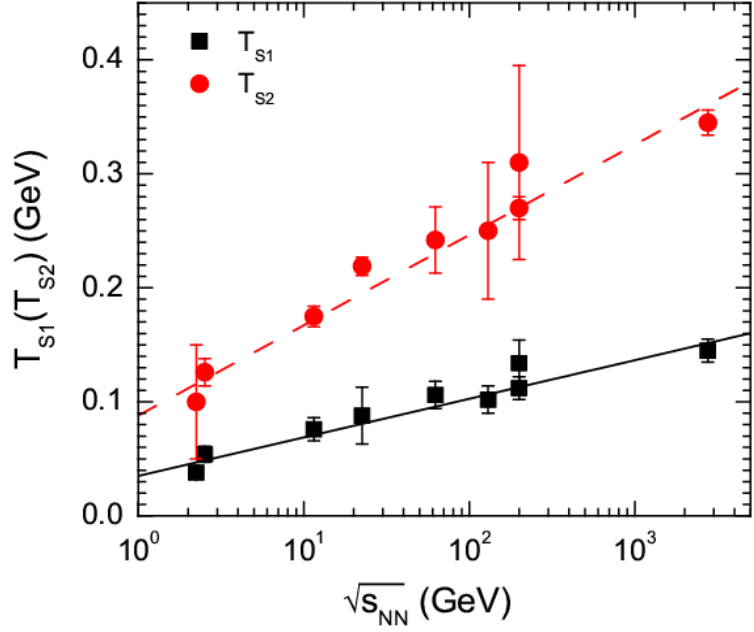

Figure 5: Dependences of $T_{S1}$ (squares) and $T_{S2}$ (circles) on $\sqrt{s_{NN}}$. The symbols represent the values of parameters taken from Table 1, and the error bars represent the statistical errors. The lines are the results fitted by the linear functions $T_{S1,S2} = a \ln \sqrt{s_{NN}} + b$.

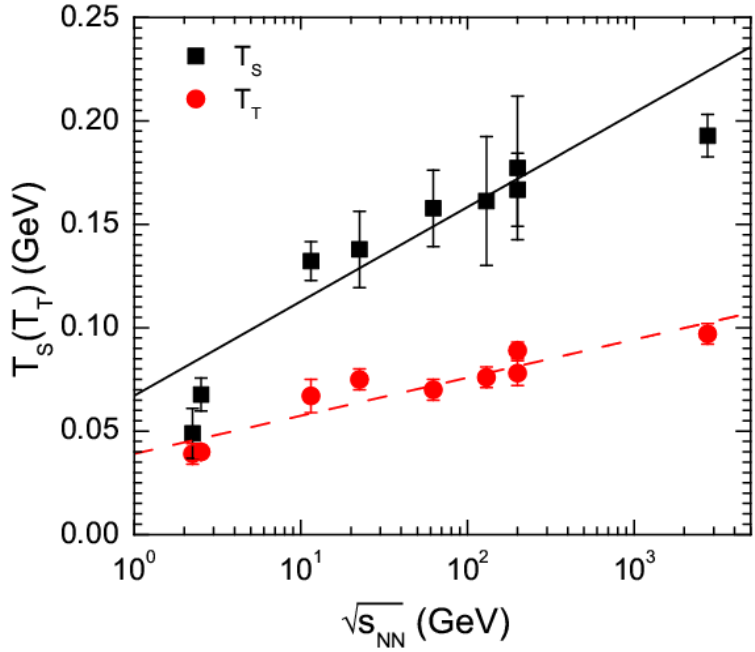

Figure 6: Same as Figure 5, but for the dependences of $T_S$ (squares) and $T_T$ (circles) on $\sqrt{s_{NN}}$ and the linear functions are $T_{S,T} = a \ln \sqrt{s_{NN}} + b$.

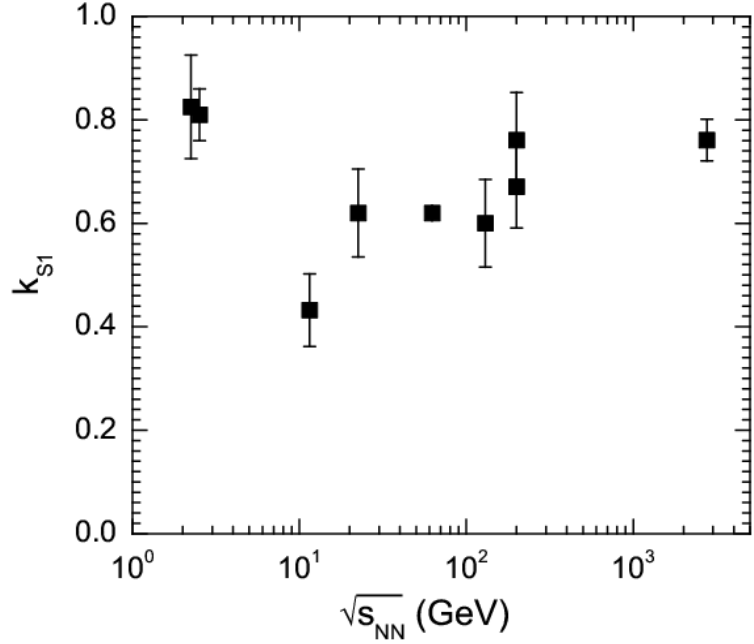

Figure 7: Dependence of $k_{S1}$ on $\sqrt{s_{NN}}$. The symbols represent the values of parameter taken from Table 1 and the error bars represent the statistical errors.

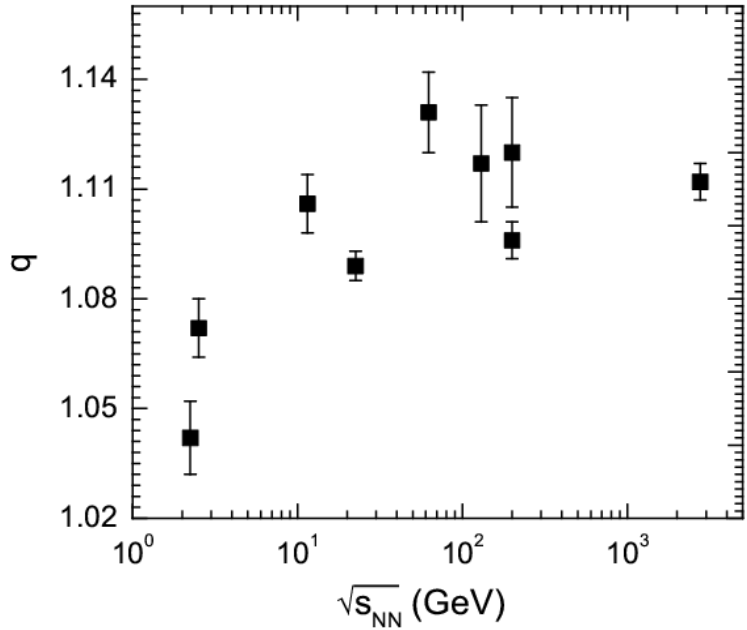

Figure 8: Dependence of $q$ on $\sqrt{s_{NN}}$. The symbols represent the values of parameter taken from Table 1 and the error bars represent the statistical errors.

Our observation that $k_{S1}$ has a minimum at about 10 GeV and $q$ increases primitively and saturates at about 10 GeV is in agreement with recent work of Cleymans [53] in which the energy region $\sqrt{s_{NN}} \approx 10$ GeV for heavy-ion collisions is indicated to be an interesting one. In fact, in this energy region, the final state has the highest net baryon density, and a transition from a baryon dominated to a meson dominated final state takes place. At the same time, rations of strange particles to mesons show obviously maxima in this energy region [53]. At a slightly smaller energy (about 6~8 GeV), other works show some extremes or saturation in excitation functions of parameters. These parameters include, but are not limited to, the specific reduced curvature of net-proton rapidity distribution [54-56], chemical freeze-out temperature [57, 58], mean transverse mass minus rest mass [57], yield rations of positive kaons to pions [57-59], squared speed-of-sound [60], string tension in Schwinger mechanism [61], width and fraction of fragmentation source [62], and width rations of experimental negative pion rapidity distribution to Landau hydrodynamic model prediction [58].

In the above analyses, for a not too wide transverse momentum spectrum, a standard distribution is usually not enough to describe the spectrum. Generally, we need a two-component standard distribution to describe the not too wide spectrum. It is expected that, in the case of studying a wider transverse momentum spectrum, we need a three-component standard distribution to describe the wider spectrum. If a set of experimental data is described by the two- or three-component standard distribution, it is also described by the Tsallis form of standard distribution [11]. If the two- or three-component standard distribution describes a temperature fluctuation between two or among three emission sources, the Tsallis form of standard distribution describes a degree of non-equilibrium. The degree of non-equilibrium is characterized by the entropy index $q$, where the larger than 1 $q$ is, the more non-equilibrium among different emission sources has. One can see from Table 1 that $q < 1.13$ in most cases, which renders an approximate equilibrium among different emission sources or the whole interacting system stays in an approximate equilibrium state.

Both the two- or three-component standard distribution and the Tsallis form of standard distribution describe only the results of soft excitation process. For the soft process, the particle spectrum appears with the characteristics of thermal emission phenomenon. Although the standard distribution describes the characteristics of thermal emission, some non-thermal emissions also obey the standard distribution. Even if the Tsallis form has less connection with thermal emission, they are relative due to the standard distribution. In the case of studying a very wide transverse momentum spectrum, for example for a width of more than 5 GeV/$c$, to consider only the contribution of soft process is not enough in description of experimental data. To describe a wider transverse momentum spectrum, we have to consider simultaneously the contribution of hard scattering process. As mentioned in section 2, According to the QCD calculus [18-20], the hard process can be described by the inverse power-law. Because of the hard process having no connection with the thermal emission, it does not affect the extraction of temperature parameter. In the case of extracting only temperature parameter, a too wide transverse momentum spectrum is not needed.

In the above analyses, the temperature extracted by us is in fact the effective temperature $T$. It is neither the temperature $T_0$ at the kinetic freeze-out nor the temperature $T_{ch}$ at the chemical freeze-out of the emission source or interacting system. Generally, $T_0$ can be extracted from the transverse momentum spectra, and $T_{ch}$ can be extracted from the ratios of different types of particles. However, the temperature extracted from the transverse momentum spectra is not sure $T_0$ due to the contribution of flow effect. How to get rid of the contribution of flow effect is a question that is worth to be discussed. In the blast-wave model [28-31], the mean transverse flow velocity $\langle \beta_T \rangle$ is introduced. Thus, $T_0$ and $\langle \beta_T \rangle$ can be simultaneously obtained based on the analysis of transverse momentum spectrum by the model. In addition, by using the standard distribution and its Tsallis form to analyze the transverse momentum spectra of different particles, we can obtain the linear relation between $T$ and $m_0$. The intercept in the linear relation is regarded as $T_0$ [21, 29, 34-36]. We can also obtain the linear relation between $\langle p_T \rangle$ and mean moving mass $\bar{m}$ (mean energy). The slope in the linear relation is regarded as $\langle \beta_T \rangle$ [37-40].

From the first and last, the two-component standard distribution and the Tsallis form of standard distribution are the same in essentials while differing in minor points in the behaviors in the figures. The standard distribution corresponds to the classical statistical system which has short-range interactions in interior and non-multi-fractal structure in boundary. Some extensive thermodynamic quantities such as energy, momentum, internal energy, entropy, and others are linearly related to the system size and particle number. These quantities obey simply additive property. The statistical method and the microscopic description of system are adaptive. The entropy function is a power tool to study the microscopic dynamics of system under the macroscopic condition by describing the occupation number of phase space of system. The Tsallis form breaks through the limitation of classical statistics by using the entropy index $q$. The complex system with long-range interactions, local memory effect, strong dynamic correlation, fractal or multi-fractal occupation in phase space, and others can be described by the Tsallis form. The Tsallis form also causes the classical extensive quantities not to obey the simple additive property. Instead, the coupled item appears in the quantities and the non-extensive statistical effects are formed in the transverse and longitudinal dynamics [63-73].

In the above discussions, one can see that the two- or three-component standard distribution can be described by the Tsallis form of stardard distribution. It does not mean that the single standard distribution cannot be described by the Tsallis form. In fact, by using a lower temperature and an entropy index that is closer to 1, the Tsallis form describes well the single standard distribution. The standard distribution is successfully replaced by the Tsallis form due to $q$ changing from 1 to a value that is greater than 1. This means that the interacting system changes from the classical and extensive statistical system to the non-extensive system, which is an essential change of the system properties. However, in some cases, the same set of experimental data can be described by both the (two- or three-cmponent) standard distribution obeyed the extensive statistics and the Tsallis form obeyed the non-extensive statistics. This means that in these cases there is no obvious boundary to distinguish extensive system and non-extensive system for a given interacting system. We have to examine which property is the main factor. Or, the interacting system in the present energy range stays in a transition gradation from extensive system to non-extensive system.

**4. Conclusions**

We summarize here our main observations and conclusions.

(a) The transverse momentum spectra of negatively charged pions produced in central nucleus-nucleus collisions measured (simulated) in mid-rapidity range by different collaborations at the SIS, RHIC, and LHC are studied by the two-component standard distribution and the Tsallis form of standard distribution which are fitted into the frame of multi-source thermal model. The two distributions describe approximately the experimental (simulated) data.

(b) The excitation functions of related parameters are analyzed. The four effective temperatures $T_{S1}$, $T_{S2}$, $T_S$, and $T_T$ increase linearly with increase of $\ln\sqrt{s_{NN}}$. In particular, the relation between $T_S$ and $T_T$ can be obtained to be $T_S = (2.500 \pm 0.170)T_T + (-0.040 \pm 0.013)$ which shows a linear relation between $T_S$ and $T_T$. With increase of $\sqrt{s_{NN}}$, $k_{S1}$ has a minimum at about 10 GeV, and $q$ increases primitively and saturates at about 10 GeV.

(c) There is no difference in the particle production in central nucleus-nucleus collisions from a few GeV to a few TeV. Combining with other works, one can say that the same or similar fits are good for proton-proton collisions. This suggests universality in particle production, as it is already obtained in mean multiplicity, pseudorapidity density, and multiplicity distribution, but now for transverse momentum distribution as well.

(d) The energy of $\sqrt{s_{NN}} \approx 10$ GeV for heavy-ion collisions is indicated to have the highest net baryon density and the maximum rations of strange particles to mesons, and to take place a transition from a baryon dominated to a meson dominated final state [53]. At a slightly smaller energy (about 6~8 GeV), other works show some extremes or saturation in excitation functions of some parameters [54-62]. These extremes and saturation are related to the search of soft point of equation-of-state.

(e) To be closer to the classical situation, the two- or three-component standard distribution has an advantage over the Tsallis form of standard distribution due to similar statistics for the classical situation and standard distribution. However, the Tsallis form of standard distribution uses less parameter than the two- or three-component standard distribution. If the two- or three-component standard distribution describes a temperature fluctuation between two or among three sources, the Tsallis form of standard distribution describes a degree of non-equilibrium.

(f) In the considered energy range, different emission sources stay in an approximate equilibrium state or the whole interacting system stays in an approximate equilibrium state. There is no obvious boundary to distinguish extensive system and non-extensive system for a given interacting system. The interacting system stays in a transition gradation from extensive system to non-extensive system. To obtain only the kinetic freeze-out temperature, we would rather use the two- or three-component standard distribution due to it being closer to the classical situation.

**Conflicts of Interest**

The authors declare that they have no conflicts of interest.

**Acknowledgments**

Comments on the manuscript and relevant communications from Edward K. G. Sarkisyan and Ya-Hui Chen are highly acknowledged. This work was supported by the National Natural Science Foundation of China under Grant Nos. 11575103 and 11747319, the Shanxi Provincial Natural Science Foundation under Grant No. 201701D121005, and the Fund for Shanxi "1331 Project" Key Subjects Construction.